\documentclass[opre,nonblindrev]{informs3_hide}

\OneAndAHalfSpacedXI 

\newcommand{\bE}{\mathbb{E}}

\newcommand{\bI}{\mathbbm{1}}
\newcommand{\eps}{\varepsilon}
\newcommand{\ALG}{\mathsf{ALG}}
\newcommand{\GREEDY}{\mathsf{Greedy}}

\newcommand{\OPT}{\mathsf{OPT}}
\newcommand{\Bern}{\mathsf{Bern}}
\newcommand{\LOST}{\mathsf{LOST}}

\newcommand{\ssA}{A^*}
\newcommand{\ssF}{F^*}
\newcommand{\sumt}{\sum_{t=1}^T}
\newcommand{\cH}{\mathcal{H}}
\newcommand{\cI}{\mathcal{I}}
\newcommand{\cC}{\mathcal{C}}
\newcommand{\ssI}{I^*}

\usepackage{natbib,bbm,multirow,multicol}
 \bibpunct[, ]{(}{)}{,}{a}{}{,}%
 \def\bibfont{\small}%

\usepackage[stable]{footmisc} 
\usepackage[normalem]{ulem}
\usepackage[dvipsnames]{xcolor}
\usepackage{float} 
\usepackage{mathtools} 
\usepackage[capitalise]{cleveref}
\usepackage{anyfontsize}

\def\blue#1{{\color{blue} #1}}
\TheoremsNumberedThrough     
\ECRepeatTheorems

\EquationsNumberedThrough    

\begin{document}


\TITLE{Leveraging Reusability: Improved Competitive Ratio of Greedy for Reusable Resources}

\ARTICLEAUTHORS{
\AUTHOR{Jackie Baek}
\AFF{NYU Stern School of Business, \EMAIL{baek@stern.nyu.edu}}
\AUTHOR{Shixin Wang}
\AFF{CUHK Business School, The Chinese University of Hong Kong, \EMAIL{shixinwang@cuhk.edu.hk}}
}
\ABSTRACT{

\OneAndAHalfSpacedXI

\footnotesize

We study online weighted bipartite matching of reusable resources where an adversarial sequence of requests for resources arrive over time. 
A resource that is matched is `used' for a random duration, drawn independently from a resource-dependent distribution, after which it returns and is able to be matched again.
We study the performance of the greedy policy, which matches requests to the resource that yields the highest reward.
Previously, it was known that the greedy policy is 1/2 competitive against a clairvoyant benchmark that knows the request sequence in advance.
In this work, we improve this result by introducing a parameter that quantifies the \textit{degree of reusability} of the resources.
Specifically, if $p$ represents the smallest probability over the usage distributions that a matched resource returns in one time step, the greedy policy achieves a competitive ratio of $1/(2-p)$.
Furthermore, when the usage distributions are geometric, we establish a stronger competitive ratio of $(1+p)/2$, which we demonstrate to be tight.
Both of these results align with the known results in the two extreme scenarios: $p=0$ corresponds to non-reusable resources, where 1/2 is known to be tight, while $p=1$ corresponds to every resource returning immediately, where greedy is the optimal policy and hence the competitive ratio is 1. 
Finally, we show that both results are robust to approximations of the greedy policy.
Our work demonstrates that the reusability of resources can enhance performance compared to the non-reusable setting, and that a simple greedy policy suffices when the degree of reusability is high. Our insights contribute to the understanding of how resource reusability can influence the performance of online algorithms, and highlight the potential for improved performance as the degree of reusability increases.

\DoubleSpacedXI

}



\maketitle

\section{Introduction}\label{s.intro}

Online bipartite matching is a classical problem where resources are matched to arriving requests that arrive one at a time \citep{karp1990optimal}.
This problem has gained significant attention in recent years due to its practical relevance in various applications, such as online marketplaces, online advertising, and resource scheduling.
Our work studies a setting where resources are \textit{reusable}, where a matched resource is unavailable for a random duration, after which they are returned and able to be matched again.
Examples of reusable resources in the real world include machines in cloud computing platforms, such as Amazon Web Services and Microsoft Azure; home and vacation rentals in online marketplaces, such as Airbnb and Vrbo; or professional services in online labor platforms such as Upwork and Fiverr.

Developing and understanding the performance of algorithms in the online matching problem with reusable resources has gained significant traction in recent years, as evidenced by the growing body of literature in this area --— see \cref{s.literature} for a comprehensive review.
In contrast to some of the recent advancements that develop \textit{new} policies to improve performance guarantees (e.g., \cite{delong2022online}),
our work studies the performance of the \textit{greedy} policy. This policy, which matches each arriving request to the resource that offers the highest immediate reward, is often regarded as the most intuitive and appealing choice for this problem. Its simplicity in both understanding and implementation makes it particularly attractive for practical applications.
Therefore, characterizing when such a policy achieves strong performance is valuable and leads to practical insights of understanding the settings in which the simple greedy policy is sufficient.
In this work, we demonstrate that the `reusability' of resources can significantly improve the performance of the greedy algorithm. By examining the interplay between resource reusability and algorithm performance, we aim to contribute to a deeper understanding of the design and evaluation of algorithms for online bipartite matching problems with reusable resources.

Formally, we consider an online bipartite matching problem with $N$ vertices on one side (`resources') and $T$ vertices on the other (`requests').
Request-side vertices arrive one at a time, and upon arrival, the resources incident to the request are revealed, and we must match the request to at most one available and adjacent resource.
When resource $i$ is matched at time $t$, we gain a reward $r_i$, and resource $i$ becomes unavailable for a random duration $D_{it} \geq 1$ that is drawn independently from a distribution $F_i$.
The resource returns at time $t+D_{it}$ and it can immediately be matched again.

We define $\GREEDY$ to be the policy that matches each request to the available resource that yields the highest immediate reward $r_i$.
We evaluate the performance of $\GREEDY$ via the competitive ratio, the worst-case ratio over instances between the expected reward of $\GREEDY$ and the expected reward of an optimal benchmark algorithm.
The optimal benchmark that we use is the optimal algorithm by a clairvoyant that knows the entire bipartite graph between requests and resources in advance, but does not know the realizations of the random usage durations.
Previously, the best known competitive ratio for $\GREEDY$ was 1/2 \citep{gong2022online} --- this matches the competitive ratio of $\GREEDY$ in the classical \textit{non-reusable} setting, where it is also known that 1/2 is tight \citep{mehta2013online}.

However, it is intuitive that resources being \textit{reusable} should improve the performance of $\GREEDY$. This improvement stems from the fact that any suboptimal matching decisions are less costly, as resources will eventually become available again for future matches. As a result, a myopic strategy such as $\GREEDY$, which focuses on maximizing the immediate reward, could potentially perform well in this scenario.
In the extreme case where each resource returns immediately within one time step, $\GREEDY$ is indeed the optimal algorithm. This is because the resources are effectively unlimited, allowing the algorithm to always match requests to the highest-reward resources without having to worry about future availability. In such a scenario, the $\GREEDY$ algorithm is perfectly suited to exploit the inherent structure of the problem.
Motivated by this intuition, we seek to characterize the performance of the $\GREEDY$ algorithm by introducing a parameter that captures the usage duration distributions of resources. This parameter effectively measures how quickly resources return to the pool of available options after being matched. By analyzing the relationship between this parameter and the performance of the $\GREEDY$ algorithm, we aim to provide a deeper understanding of the conditions under which the simple greedy policy performs well. Ultimately, our analysis will offer valuable insights into the design and evaluation of algorithms for online bipartite matching problems with reusable resources.

\textbf{Our Results.}
Letting $p = \min_i \Pr(F_i = 1)$, the smallest probability over all resources that a matched resource returns in the next time step, 
we show that $\GREEDY$ achieves a competitive ratio of $\frac{1}{2-p}$.
Distributions with a hazard rate of at least $p$ fit this assumption.
Our guarantee aligns with existing findings in two extreme cases:
When $p = 0$, resources are not reusable, corresponding to the classical setting where the tight competitive ratio is $1/2$.
When $p = 1$, each resource returns immediately, rendering the $\GREEDY$ algorithm optimal and yielding a competitive ratio of 1.
Next, in the special case where all usage duration distributions are geometric with parameter at least $p$, we prove a stronger competitive ratio of $\frac{1+p}{2}$.
This ratio matches the first result at the endpoints $p=0$ and $p=1$, but offers a strictly better guarantee for $p \in (0, 1)$.
This second result is tight, in that there is a simple instance with two time steps where $\GREEDY$ achieves a $\frac{1+p}{2} + \eps$ fraction of the reward as the clairvoyant for any $\eps > 0$.
Lastly, we show that both results are robust to approximations of the $\GREEDY$ policy. 
Specifically, if an algorithm matches requests to resources whose reward is at least a factor $\alpha \in [0, 1]$ of the reward of the resource that $\GREEDY$ would match,
then the competitive ratio of this algorithm under the two assumptions are $\frac{1}{1-p + 1/\alpha}$ and $\frac{1}{(1-p)/(1+p) + 1/\alpha }$ respectively.

\textbf{Proof Idea.}
We use a sample path coupling technique that extends the proof from \cite{gong2022online} which demonstrates the 1/2 competitive ratio.
Letting $\OPT$ be the optimal algorithm by the clairvoyant, we define a notion of a `lost match' to be an event where $\OPT$ matched a resource $i$ at a time when $i$ was unavailable under $\GREEDY$.
The crux of our analysis relies on upper bounding the probability of a lost match, and we do this in two ways.
The first way relies on the reusability of resources --- if $p$ is the minimum probability of return, every resource will be available at each time step with a probability of at least $p$.
Next, the second way relies on characterizing and studying an event that would prevent a lost match from occurring.
Specifically, a lost match at time $t$ occurs because $\GREEDY$ matched the resource at an earlier time $t' < t$, and it has not returned by $t$. Then, if $\OPT$ had also matched the same resource at time $t'$, and the usage durations between the algorithms were coupled, a lost match cannot happen at time $t$.
We analyze the probability of this event and relate it to the event of a lost match.
This step uses an intricate set of arguments that rely on the usage duration distributions to be geometric, and hence this step is used for the second result with the geometric assumption.

Prior to our work, the performance of online algorithms for reusable resources was worse than or equal to the performance guarantees for {non-reusable} resources, due to the analytical difficulties caused by the reusability. 
In contrast, our work \textit{leverages reusability} to show that the $\GREEDY$ policy can achieve a better performance guarantee compared to the non-reusable setting.
The main insight behind this phenomenon is that the reusability of resources can mitigate the future loss caused by a suboptimal match, since the resource will return in the future and can be matched again.
This leads to the managerial insight that when the degree of reusability is high, a simple greedy policy will perform well.
To the best of our knowledge, we are the first to characterize how the performance of an online algorithm changes with the degree of resource reusability.

\subsection{Related Literature} \label{s.literature}
We first briefly describe the literature on the classical online matching problem with non-reusable resources.
In the setting where the request sequence is adversarial, the greedy policy achieves a 1/2 competitive ratio \citep{mehta2013online}, while a randomized ranking policy is shown to be optimal with a competitive ratio of $1-1/e$ \citep{karp1990optimal,birnbaum2008line, goel2008online}. \cite{aggarwal2011online} introduced the online vertex-weighted bipartite matching problem, and show that the randomized Perturbed Greedy algorithm, which is a strict generalization of the ranking algorithm, can also achieve the $1-1/e$ competitive ratio. When each resource can have a high inventory, \cite{kalyanasundaram2000optimal} propose a deterministic Balance algorithm that achieves an $1-1/e$ competitive ratio.
\cite{mehta2007adwords} introduce the AdWords problem and propose a novel algorithm that achieves an $1-1/e$ competitive ratio, which is a generalization of Ranking and Balance.
\cite{buchbinder2007online,devanur2013randomized} provide an elegant proof of a general version of the above algorithms from the randomized primal-dual view.
We refer the reader to \cite{mehta2013online} for a more detailed discussion.
Online decision making has also been studied in pricing \citep{ma2020algorithms}, network revenue management \citep{ma2020approximation}, assortment optimization \citep{chan2009stochastic,golrezaei2014real} and online stochastic knapsack \citep{jiang2022tight}. 
In all the settings mentioned above, no deterministic algorithm can achieve a competitive ratio greater than $\frac{1}{2}$ without the large-inventory assumption. 

Incorporating the reusability of resources brings new challenges to the analysis of online matching algorithms.
The existing literature has aimed to achieve the same performance guarantees in the reusable setting as compared to the non-reusable setting.
The adversarial arrival model, which we focus on in this work, was first studied in \cite{gong2022online}, where it was shown that the greedy algorithm is $1/2$ competitive.
\cite{delong2022online} provide a 0.589 competitive policy that is a generalization of the classic RANKING algorithm for non-reusable resources.
With the additional assumption of large inventory, \cite{goyal2020asymptotically} and \cite{feng2021online} propose different algorithms that achieve a $1-1/e$ competitive ratio, under the online matching problem and online assortment problem, respectively. \cite{huo2022online,zhang2022online} study the online resource allocation problems with demand-dependent rewards.

Reusable resources have also been studied under stochastic arrivals, where the distribution of the request arrivals is known.
\cite{dickerson2021allocation} provide a simulation-based algorithm that achieves a $1/2$ competitive ratio for the online matching problem. 
\cite{rusmevichientong2020dynamic} studies the problem in an assortment setting, and showed a 1/2 approximation compared to the optimal DP based on approximate dynamic programming.
\cite{feng2019linear} and \cite{baek2022bifurcating} proposes different algorithms that are 1/2 competitive against the LP benchmark.
\cite{feng2022near} propose a near-optimal algorithm under the infinite inventory regime. 
\cite{rusmevichientong2023revenue} study the resource allocation problem in the hotel room booking setting and propose algorithms whose performance depends on the maximum usage duration.
\cite{xiedynamic} study the asymptotic regime with many resource units in an overloaded network and propose an algorithm that achieves a logarithmic regret.

Besides the online matching setting, there are also works considering the pricing problem of reusable resources. \cite{besbes2019static} study the pricing problem with reusable resources, and they prove that the static pricing obtains a $78.9\%$ performance guarantee simultaneously for profit, market share and service level from optimal policy. \cite{jia2022online} study the online pricing problem for reusable resources where the service rate and service rate are exponential and price-dependent. \cite{banerjee2022pricing} study the pricing decision in a vehicle-sharing system and provide an algorithm that achieves a performance guarantee that only depends on the average units per station.

\section{Model and Results}\label{s.model}

We consider an online matching problem where $G$ is a bipartite graph with $N$ vertices on the `offline' side, $T$ vertices on the `online' side, and edges $E$ denoting whether the offline vertex can be used to match the online vertex.
We refer to the vertices on the offline side as `resources', and vertices on the online sides as `requests'.
At each time step $t=1, \dots, T$, the online vertex $t$ arrives, the edges incident to $t$ are revealed, and we must make an irrevocable decision to match $t$ to at most one resource that is not already matched.
If we match $t$ to resource $i$, then $i$ is unavailable for a random duration $D_{it} \geq 1$, which we refer to as the `usage duration'.
Resource $i$ returns at the start of time $t + D_{it}$ and is immediately available to be matched again.
For each resource $i$, we assume that there is a distribution $F_i$ in which the usage durations are drawn from independently whenever $i$ is matched.
We earn a reward of $r_i$ whenever resource $i$ is matched, and we assume that $0 \leq r_1 \leq \dots \leq r_N$.

Let $I_t \subseteq [N]$ denote the set of resources that are available to be matched at time $t$, and let $N_t \subseteq [N]$ be the set of resources incident to request $t$.
We denote by $A_t \in (N_t \cap I_t) \cup \{0\}$ as the resource that is matched at time $t$, where $A_t = 0$ represents the no-match action.
The $\GREEDY$ policy matches the available resource with the highest reward; i.e. $A_t = \max ((N_t \cap I_t) \cup \{0\})$.
We compare the performance of $\GREEDY$ to an optimal clairvoyant benchmark that knows the entire arrival sequence but does not know the realizations of the usage durations $D_{it}$. We refer to this clairvoyant by $\OPT$.

A problem instance $\cI$ is determined by the bipartite graph $G$, reward $r_i$ and usage duration distribution $F_i$ for all offline vertices $i\in N$, i.e. $\cI = (G, \{r_i, F_i\}_{i \in [N]})$. Let $\OPT(\cI)$ and $\GREEDY(\cI)$ be the expected reward of $\OPT$ and $\GREEDY$ respectively.
We do not specify the dependence on $\cI$ when it is clear.
For a class of instances $\cC$, we say that $\GREEDY$ achieves a competitive ratio of $c$ if for every instance $\cI \in \cC$, $\frac{\GREEDY(\cI)}{\OPT(\cI)} \geq c$.
We analyze the competitive ratio for $\GREEDY$ under two different assumptions of usage durations.


\subsection{Results}
First, we prove an upper bound for $\frac{\GREEDY(\cI)}{\OPT(\cI)}$ that is parameterized by $p = \min_i \Pr(F_{i} = 1)$, the smallest probability over all resources that the resource returns immediately in the next time step.

\begin{theorem} \label{theorem:weak}
Given an instance $\cI$, let $p = \min_i \Pr(F_{i} = 1)$. 
Then,
$$\frac{\GREEDY(\cI)}{\OPT(\cI)} \geq \frac{1}{2-p}.$$
\end{theorem}
\cref{theorem:weak} implies that for any fixed $p \in [0, 1]$, for the class of instances that satisfy $\min_i \Pr(F_{i} = 1) \geq p$, 
$\GREEDY$ achieves a competitive ratio of $1/(2-p)$.

Next, we consider the class of instances where the usage durations are geometric.
\begin{assumption} \label{assump:geometric}
$F_{i} = \text{Geometric}(p_i)$ for some $p_i \geq 0$ for all $i \in [N]$.
\end{assumption}

\begin{theorem} \label{theorem:main}
For any instance $\cI$ that satisfies \cref{assump:geometric}, if $p=\min_i p_i$, then
$$\frac{\GREEDY(\cI)}{\OPT(\cI)} \geq \frac{1+p}{2}.$$
\end{theorem}

Under both Theorem \ref{theorem:weak} and Theorem \ref{theorem:main}, the competitive ratio begins at 1/2 when $p = 0$ and monotonically increases to 1 as $p$ increases to 1. The case where $p = 0$ corresponds to non-reusable resources, and both results are consistent with the 1/2 competitive ratio for online matching as established by \citep{mehta2013online}. On the other hand, when $p = 1$, any matched resource returns immediately in the next time step, making the $\GREEDY$ policy the optimal solution.

For instances with $p \in (0, 1)$, the bound provided by Theorem \ref{theorem:main} is strictly stronger than the one given by Theorem \ref{theorem:weak}. We will now illustrate an example demonstrating that Theorem \ref{theorem:main} is tight. In other words, for any $p$ and $\epsilon > 0$, there exists an instance satisfying Assumption \ref{assump:geometric} such that $\frac{\GREEDY(\cI)}{\OPT(\cI)} \leq \frac{1+p}{2} + \epsilon$.

\begin{example}
Fix $p \in [0, 1]$.
Let $T = 2$, $N=2$ and $\delta > 0$. Let $r_1 =1, r_2 = 1 + \delta$, and let $N_1 = \{1, 2\}, N_2 = \{2\}$. Suppose $p_1 = p_2 = p$.
The clairvoyant will choose item 1 at time 1 and item 2 at time 2, which yields a reward of $\OPT = 2+\delta$.
$\GREEDY$ will choose item 2 at time 1. If item 2 returns at time 2, $\GREEDY$ will choose item 2, otherwise there will be no allocation.
The expected reward is $\GREEDY = (1+p) (1+\delta)$.
Then, $\GREEDY/\OPT = (1+p) (1+\delta)/(2+\delta) \to (1+p)/2$ as $\delta \to 0$.
\end{example}

Lastly, we consider an algorithm $\ALG$ that is not exactly the $\GREEDY$ policy, but approximates it in terms of the rewards of the resources that it matches.
We show in the next theorem that $\ALG$ also admits a competitive ratio that scales gracefully with how well $\ALG$ approximates $\GREEDY$.
To facilitate our analysis, let us define $r^*(I) = \max_{i \in I} r_i$ for a set of resources $I \in [N]$, representing the highest reward within the given set.

\begin{theorem} \label{thm:approx}
Let $\ALG$ be a policy such that there is an $\alpha \in [0, 1]$ such that
\begin{align} \label{eq:approx}
\bE\left[ \sumt  r_{A_t}  \right] 
\geq \alpha \bE\left[ \sumt  r^*(I_t) \right],
\end{align}	
where $A_t$ is the resource matched at time $t$ by $\ALG$, and $I_t$ are the resources available under $\ALG$ at time $t$.
Then, given an instance $\cI$, for $p = \min_i \Pr(F_i = 1)$,
\begin{align*}
\frac{\ALG(\cI)}{\OPT(\cI)} \geq \frac{1}{1-p + 1/\alpha}.
\end{align*}
Next, if the usage durations are geometric and $p = \min p_i$, 
\begin{align*}
\frac{\ALG(\cI)}{\OPT(\cI)} \geq \frac{1}{(1-p)/(1+p) + 1/\alpha}.
\end{align*}
\end{theorem}

For example, if $\ALG$ always chooses a resource with a reward that is at least half of that of the highest reward resource, then \eqref{eq:approx} is satisfied with $\alpha = 1/2$, which results in an approximation of $\frac{1}{3-p}$ and $\frac{1+p}{3+p}$ under the two assumptions respectively. 
The condition \eqref{eq:approx} is weaker than $\ALG$ needing to satisfy the approximation at \textit{every} time step; rather the total approximation needs to be satisfied in expectation over all time steps.
This result is useful in settings where the exact reward is not known or needs to be estimated, or in other situations where business constraints prevent the implementation of the precise greedy policy. By demonstrating that the competitive ratio scales gracefully with the level of approximation, $\alpha$, our findings reveal that even a suboptimal approximation of the $\GREEDY$ policy can achieve strong performance in specific settings. This insight can be particularly useful for practitioners who must navigate the complexities of real-world environments where perfect information or adherence to an ideal policy may not be feasible.

\section{Proofs} \label{s.proof}

The proofs of \cref{theorem:weak} and \cref{theorem:main} start with the same steps.
Both proofs rely on coupling sample paths between $\GREEDY$ and $\OPT$, and the two results use two different coupling mechanisms, which we specify later.

For a fixed instance $\cI$, let's assume that the sample paths of $\GREEDY$ and $\OPT$ are coupled (coupling specified later). We denote by $I_t, \ssI_t \subseteq [N]$ to be the resources that are available at time $t$ under $\GREEDY$ and $\OPT$ respectively.
Similarly, denote by $A_t \in I_t, \ssA_t \in \ssI_t$ to be the resource matched by $\GREEDY$ and $\OPT$ at time $t$ respectively.
We define $O_{it} = \{ j \notin I_t \;\forall j \geq i\}$ as the event that at time $t$, all resources $j \geq i$ are \textit{unavailable} under $\GREEDY$.

For each resource $i$ matched under $\OPT$, we decompose the reward based on whether $O_{it}$ occurs:
\begin{align}
\OPT 
&= \bE\left[ \sumt \sum_{i \in [N]} r_i \bI( \ssA_t = i)\right] \nonumber \\
&= \bE\left[ \sumt \sum_{i \in [N]} r_i \bI( \ssA_t = i, \neg O_{it}) \right] 
 +  \bE\left[ \sumt \sum_{i \in [N]} r_i \bI( \ssA_t = i, O_{it})\right] \label{eq:decomp1} 
\end{align}
When $\neg O_{it}$ occurs, then $\GREEDY$ will match a resource with higher reward than resource $i$ at time $t$ by definition of greedy.
Therefore, the first term of \eqref{eq:decomp1} is at most $\GREEDY$.
Denote the second term in \eqref{eq:decomp1} by $\LOST = \bE\left[ \sumt \sum_{i \in [N]} r_i \bI( \ssA_t = i, O_{it})\right]$, so that we have
\begin{align*}
\OPT \leq \GREEDY  + \LOST.
\end{align*}
$\LOST$ represents the reward gained when $\OPT$ matches a resource when $\GREEDY$ did not have that resource (or more valuable) available, hence we refer to this as a `lost' matching.
Subsequently, we present two propositions that upper bound $\LOST$ under the two different assumptions, which lead to Theorem \ref{theorem:weak} and Theorem \ref{theorem:main}, respectively.
\begin{proposition} \label{prop:bound_lost_weak}
If $p = \min_i \Pr(D_{i1} = 1) \geq 0$,
\begin{align*}
\LOST \leq (1-p)\GREEDY.
\end{align*}
\end{proposition}

\begin{proposition} \label{prop:bound_lost}
Under \cref{assump:geometric},
\begin{align*}
\LOST \leq \frac{1-p}{1+p}\GREEDY.
\end{align*}
\end{proposition}

We prove these two propositions in the following subsections.
We note that the proof of \cref{prop:bound_lost} builds off of the ideas from the proof of \cref{prop:bound_lost_weak}.

\subsection{Proof of \cref{prop:bound_lost_weak}}

For this proof, we use a similar sample path coupling to one that was used in \cite{gong2022online}, which enforces that $\OPT$ have the same usage duration realizations as $\GREEDY$.

\paragraph{Sample path coupling.}
For each resource $i$, we maintain a stack $S_i$, which is initially empty.
At a time step $t$, if both $\GREEDY$ and $\OPT$ match the same resource $i$, then we generate an i.i.d. sample from $F_i$ and use this as the usage duration for both policies.
Otherwise, when a resource $i$ is matched under $\GREEDY$, we generate an i.i.d. sample from the distribution $F_i$ which we push onto the stack $S_i$, as well as use for resource $i$'s usage duration under $\GREEDY$.
When a resource $i$ is matched under $\OPT$, we pop from the stack $S_i$ (LIFO order) and use that as the usage duration, which we remove from the stack.
If the stack was empty, then we simply generate an i.i.d. sample from $F_i$. This sample path coupling method is designed to synchronize the usage durations of resources for both $\GREEDY$ and $\OPT$ algorithms, allowing for a fair comparison of their performance. It ensures that the usage duration realizations are shared between the two policies whenever possible, preserving the dependencies between the realizations while still accounting for the differences in the matching decisions made by each algorithm.



\proof{Proof of \cref{prop:bound_lost_weak}.}
Recall that $\LOST = \bE\left[ \sumt \sum_{i \in [N]} r_i \bI( \ssA_t = i, O_{it})\right]$.
Fix some $i$ and $t$ where the event $\{\ssA_t = i, O_{it}\}$ occurs --- we refer to this event as a `lost' match.
A lost match implies that $\GREEDY$ matched resource $i$ at an earlier time step and the resource has not returned since then.
Define $\tau_i(t) < t$ be the last time that $i$ was matched during $\GREEDY$:
\begin{align} \label{eq:deftau}
\tau_i(t) \triangleq \max\{t' < t: A_{t'} = i \text{ or } t' = 0\}.
\end{align}

Next, we claim that if lost matches occurs under times $t$ and $s$, then $\tau_i(t) \neq \tau_i(s)$.
Specifically, suppose $t < s$ such that $\{\ssA_t = i, O_{it}\}$ and $\{\ssA_s = i, O_{is}\}$.
By the sample path coupling, when $\OPT$ matches resource $i$ at time $t$, it will use the same usage duration as when $\GREEDY$ matched resource $i$ at time $\tau_i(t)$.
Therefore, the resource returns under $\GREEDY$ before it is returned under $\OPT$.
At time $s > t$, the resource $i$ is matched under $\OPT$ again; then, for $O_{it}$ to occur, it must have been that $\GREEDY$ also matched the resource again after $\tau_i(t)$. Therefore, $\tau_i(t) < \tau_i(s)$.

This one-to-one relationship of $\tau_i(\cdot)$ allows us to switch the sum of sales from $\OPT$ to sales from $\GREEDY$.
\begin{align*}
\LOST 
&= \bE\left[ \sumt \sum_{i \in [N]} r_i \bI( \ssA_t = i, O_{it})\right] \\
&= \bE\left[ \sum_{t'=1}^T \sum_{i \in [N]} r_i \bI( A_{t'} = i, \exists t > t' \text{ s.t. } i \notin I_t, \ssA_t = i, t' = \tau_i(t)) \right].
\end{align*}
Note that for the event $\{A_{t'} = i, \exists t > t' \text{ s.t. } i \notin I_t, t' = \tau_i(t)\}$ to occur, it must be that $D_{it'} > 1$; i.e. the usage duration of $i$ is longer than 1 for it to have caused a future lost match.
Therefore,
\begin{align*}
\LOST 
&\leq \bE\left[ \sum_{t'=1}^T \sum_{i \in [N]} r_i \bI( A_{t'} = i, D_{it'} > 1)\right] \\
&= \sum_{t'=1}^T \sum_{i \in [N]} r_i \Pr( A_{t'} = i ) \Pr( D_{it'} > 1 \;|\; A_{t'} = i).
\end{align*}
By the definition of $p$, $\Pr( D_{it'} > 1 \;|\; A_{t'} = i) \leq (1-p)$, which yields the desired bound of $\LOST \leq (1-p)\GREEDY$.
\Halmos
\endproof

\subsection{Proof of \cref{prop:bound_lost}}
For this proof, we use the following mechanism to couple sample paths, which makes use of the assumption of the geometric usage duration distribution.

\paragraph{Sample path coupling.}
Let $P_{it} \sim \Bern(p_i)$, drawn independently for every $i \in [N]$ and $t \in [T]$. A resource $i$ that was unavailable returns at the start of time $t$ if and only if $P_{it} = 1$. Both $\GREEDY$ and $\OPT$ share the same $P_{it}$ random variables.

Now, we build off of the proof of \cref{prop:bound_lost_weak} to tighten the bound under the stronger assumption of \cref{assump:geometric}.
In the proof of \cref{prop:bound_lost_weak}, the main idea was that if a resource matched under $\GREEDY$ at time $t'$ returns immediately in the next time step (which happens with probability $\geq p$), then it would not contribute to $\LOST$.
Then, our idea is to incorporate the fact an allocation from $\GREEDY$ at time $t'$ would also not contribute to $\LOST$ if $\OPT$ matched the \textit{same} resource at time $t'$.
That is, if both $\GREEDY$ and $\OPT$ match resource $i$ at time $t'$, the resource will \textit{return at the same time} under both policies (due to the sample path coupling).
Therefore, this cannot cause the event $\{\ssA_t = i, O_{it}\}$ to happen in the future.
Hence when $\GREEDY$ matches resource $i$ at time $t'$, it would not contribute to $\LOST$ if \textit{either} the resource returns before the next time step, or if $\OPT$ had also matched resource $i$ at time $t'$.
The above arguments yield the following bound.
\begin{proposition} \label{prop:lost1}
\begin{align}
\LOST 
&\leq (1-p) \left( \GREEDY -   \sum_{t=1}^T \sum_{i \in [N]} r_i  \Pr(\ssA_{t} = A_{t} = i)  \right). \label{eq:sketchlost2}
\end{align}
\end{proposition}
The RHS of \eqref{eq:sketchlost2} tracks how much of the reward of $\GREEDY$ could contribute to $\LOST$.
A reward cannot contribute to $\LOST$ if the resource returns in the next time step (hence the $(1-p)$ term), nor if $\OPT$ matched the same resource (hence subtracted term).
\cref{prop:lost1} is formally derived in \cref{s.pfprop}.
Then, our goal is to lower bound the term $\sum_{t=1}^T \sum_{i \in [N]} r_i  \Pr(\ssA_{t} = A_{t} = i)$.
We do this via the following lemma:
\begin{lemma} \label{lemma:keylemma}
For any $i \in [N]$ and $t \in [T]$,
\begin{align*} %
\Pr(\ssA_t = A_t = i) \geq \frac{p}{1-p} \Pr(\ssA_t = i, O_{it}).
\end{align*}
\end{lemma}
This is a key result in our proof, and it relies on the usage durations to be geometric --- the proof can be found in  \cref{s.pflemma}.
The main intuition is that the difference between whether the event $\{A_t = i\}$ occurs or $O_{it}$ occurs effectively hinges on whether resource $i$ returns at the start of time $t$ or does not return, which happens with probability $p$ and $(1-p)$ respectively.
Therefore, the ratio $\Pr(\ssA_t = A_t = i)/\Pr(\ssA_t = i, O_{it})$ equals $p/(1-p)$.

Plugging \cref{lemma:keylemma} into \eqref{eq:sketchlost2} yields
\begin{align*}
\LOST 
&\leq 
(1-p) \left( \GREEDY - \frac{p}{1-p}  \sum_{t=1}^T \sum_{i \in [N]} r_i \Pr(\ssA_t = i, O_{it})  \right)  \\
&= (1-p) \left( \GREEDY - \frac{p}{1-p}  \LOST \right).
\end{align*}
Rearranging leads to the desired result of $\LOST \leq \frac{1-p}{1+p} \cdot \GREEDY$, finishing the proof of \cref{prop:bound_lost} as well as \cref{theorem:main}.

\subsection{Proof of \cref{prop:lost1}} \label{s.pfprop}
Define $\tau_i(t) < t$ be the last time that $i$ was matched during $\GREEDY$:
\begin{align*}
\tau_i(t) \triangleq \max\{t' < t: A_{t'} = i \text{ or } t' = 0\}.
\end{align*}
If the event $\{\ssA_t = i, O_{it}\}$ occurs, $\tau_i(t)$ is the time that $\GREEDY$ matched resource $i$, and resource $i$ has not returned since then.
That is, the sale at time $\tau_i(t)$ causes the lost allocation $\{\ssA_t = i, O_{it}\}$.
\begin{align}
\LOST 
&= \bE\left[ \sumt \sum_{i \in [N]} r_i \bI( \ssA_t = i, O_{it})\right] \nonumber \\
&\leq \bE\left[ \sumt \sum_{i \in [N]} r_i \bI( \ssA_t = i, i \notin I_t, A_{\tau_i(t)} = i)\right] \label{eq:1}
\end{align}
We show that for every term in \eqref{eq:1} maps to a unique $\tau_i(t)$, which holds due to the sample path coupling.
\begin{claim} \label{claim:1-1}
If $t \neq s$ such that the both events $\{\ssA_t = i, i \notin I_t\}$ and $\{\ssA_s = i, i \notin I_s\}$ occur, then $\tau_i(t) \neq \tau_i(s)$.
\end{claim}
This allows us to switch the summation to sum over $t'$:
\begin{align}
\LOST 
&\leq \bE\left[ \sum_{t'=1}^T \sum_{i \in [N]} r_i \bI( A_{t'} = i, \exists t > t' \text{ s.t. } \ssA_t = i, i \notin I_t, t' = \tau_i(t)) \right]. \label{eq:2}
\end{align}
The right hand side of \eqref{eq:2} sums over the lost rewards generated by the resources that are previously matched by $\GREEDY$, and the indicator represents whether this causes a future lost sale.

For every $i$ and $t$, define $D_{it} = \max \{d \geq 1\;:\; P_{i,t+d'}=0 \;\forall d' \leq d\}$ to be the \textit{duration} that resource $i$ would be unavailable if it was matched at time $t$.

We show that for event $\{A_{t'} = i, \exists t > t' \text{ s.t. } \ssA_t = i, i \notin I_t, t' = \tau_i(t)\}$ to happen, it must be that $\GREEDY$ and $\OPT$ matched different resources at time $t'$, and that $D_{it'} > 1$.
\begin{claim} \label{lemma:same_resource}
For any $i \in [N], t' \in [T]$, 
if $\{A_{t'} = i, \exists t > t' \text{ s.t. } \ssA_t = i, i \notin I_t, t' = \tau_i(t)\}$ occurs, then 
$\{A_{t'} = i, \ssA_{t'} \neq A_{t'}, D_{it'} > 1 \}$ occurs.
\end{claim}
This result holds due to the sample path coupling.
If both $\GREEDY$ and $\OPT$ matches resource $i$ at time $t'$, the resources will come back at the same time, and hence the event $\{A_{t'} = i, \exists t > t' \text{ s.t. } \ssA_t = i, i \notin I_t, t' = \tau_i(t)\}$ cannot occur. 
Using \cref{lemma:same_resource} yields
\begin{align*}
\LOST 
&\leq \bE\left[ \sum_{t=1}^T \sum_{i \in [N]} r_i \bI(A_{t} = i, \ssA_{t} \neq A_{t}, D_{it} > 1)  \right] \\
&= \sum_{t=1}^T \sum_{i \in [N]} r_i \Pr(A_{t} = i, \ssA_{t} \neq A_{t}) \Pr(D_{it} > 1 \;|\; A_{t} = i, \ssA_{t} \neq A_{t}).
\end{align*}
Note that $D_{it}$ is only a function of the variables $(P_{is})_{s > t}$, which are independent of $A_t$ and $\ssA_t$.
Therefore, $\Pr(D_{it} > 1 \;|\; A_{t} = i, \ssA_{t} \neq A_{t}) \leq 1-p$.
Then we have
\begin{align*}
\LOST
&\leq (1-p) \sum_{t=1}^T \sum_{i \in [N]} r_i \Pr(A_{t} = i, \ssA_{t} \neq A_{t}) \\
&= (1-p) \bE\left[ \sum_{t=1}^T \sum_{i \in [N]} r_i \bI(A_{t} = i, \ssA_{t} \neq A_{t}) \right] \\
&= (1-p) \bE\left[ \sum_{t=1}^T \sum_{i \in [N]} r_i \left( \bI(A_{t} = i) - \bI(\ssA_{t} = A_{t} = i) \right) \right] \\
&= (1-p) \left( \GREEDY -  \bE\left[ \sum_{t=1}^T \sum_{i \in [N]} r_i  \bI(\ssA_{t} = A_{t} = i) \right] \right),
\end{align*}
as desired.

\subsubsection{Proof of \cref{claim:1-1}.}

Let $t < s$ such that the both events $\{\ssA_t = i, i \notin I_t\}$ and $\{\ssA_s = i, i \notin I_s\}$ occur.
For both $\ssA_t = i$ and $\ssA_s = i$ to occur, resource $i$ must have returned in between time $t$ and $s$. That is, there exists a $t' \in \{t+1, \dots, s\}$ such that $P_{it'} = 1$.
Then, due to the sample path coupling, resource $i$ is available under $\GREEDY$ at time $t'$; i.e. $i \in I_{t'}$.
If $i \notin I_s$, then it must be that $\GREEDY$ matched resource $i$ between time $t'$ and $s$.
Then, by definition $\tau_i(s) \geq t' > t > \tau_i(t)$.

\subsubsection{Proof of \cref{lemma:same_resource}.}
Fix $i$, $t'$ such that $\{A_{t'} = i, \exists t > t' \text{ s.t. } \ssA_t = i, i \notin I_t, t' = \tau_i(t)\}$ occurs.
Suppose, to the contrary, that $\ssA_{t'} = A_{t'}$.
Then, due to the sample path coupling, resource $i$ returns at the same time $s > t'$ under $\GREEDY$ and $\OPT$.
It must be that $t \geq s$.
If $i \notin I_t$, it must be that $\GREEDY$ matched resource $i$ again before time $t$, and hence by definition, $\tau_i(t) \neq t'$.

If $D_{it'} = 1$, then $i$ is available at time $t'+1$.
Therefore, for the same reasoning, for $i \notin I_t$ to occur, $\GREEDY$ must have matched resource $i$ again, in which case $\tau_i(t) \neq t'$.

\subsection{Proof of \cref{lemma:keylemma}} \label{s.pflemma}

Recall that $O_{it} = \{ j \notin I_t \;\forall j \geq i\}$.
Then, by definition of $\GREEDY$, $O_{it}$ happens if and only if $A_t < i$.
Therefore, we would like to show
\begin{align*} %
\Pr(\ssA_t = A_t = i) \geq \frac{p}{1-p} \Pr(\ssA_t = i, A_t < i).
\end{align*}

Fix any $i$ and $t$ such that $i \in N(t)$.
Denote by $H_t = \{(P_{is})_{i \in [N]}, N_s, A_s, A^*_s\}_{s \leq t}$ the history of events of both $\GREEDY$ and $\OPT$ at the end of time $t$.

Let $F_{it} = \{i \notin I_{t}\} \cup \{A_{t} = i\}$ be the event where $i$ was either unavailable at time $t$ or matched at time $t$ under $\GREEDY$.
$F_{it}$ implies that resource $i$ is unavailable at the \textit{end} of time $t$.
Define $\ssF_{it}= \{i \notin I^*_{t}\} \cup \{A^*_{t} = i\}$ analogously for $\OPT$.

\begin{claim}
If the event $\{\ssA_t = i, A_t < i\}$ occurs, it must be that $F_{i, t-1} \cap \neg {\ssF}_{i, t-1}$ occurs.
\end{claim}

\proof{Proof.}
Suppose $\{\ssA_t = i, A_t < i\}$ is true. 
If $F_{i, t-1}$ did not happen, then resource $i$ is available at time $t$ under $\GREEDY$, and hence the greedy algorithm will match resource $i$ (or more valuable), which contradicts $A_t < i$.
Hence it must be that $F_{i, t-1}$ occurs.
Next, suppose $\ssF_{i, t-1}$ occurred.
But since $\ssA_t = i$ happens, it must be that resource $i$ returned exactly at time $t$, hence $P_{it} = 1$.
But that means resource $i$ will also be available under $\GREEDY$, and hence it cannot be the case that $A_t < i$.
Therefore, it must be that $F_{i, t-1}(i)$ occurred and $\ssF_{i, t-1}$ did not.
\Halmos
\endproof

Let $\cH_t$ be the set of histories where $F_{it} \cap \neg {\ssF}_{i, t-1}$ occurs.
For the event $A_t < i$ to occur, it must be that all of the following occur: $H_t \in \cH_t$, $P_{it} = 0$, and that $j \notin I_t \;\forall j > i$.
Then, for any history $h \in \mathcal{H}_t$,
\begin{align*}
&\Pr(\ssA_t = i, A_t < i, H_t = h)  \\
=& \Pr(\ssA_t = i, H_t = h, P_{it} = 0, j \notin I_t \;\forall j > i) \\
=& \Pr(P_{it} = 0 \;|\; \ssA_t = i, H_t = h, j \notin I_t \;\forall j > i)\Pr(\ssA_t = i, H_t = h, j \notin I_t \;\forall j > i) \\
=& (1-p_i) \Pr(\ssA_t = i, H_t = h, j \notin I_t \;\forall j > i) 
\end{align*}
The last equality holds because $P_{it}$ is independent of all of the events in the conditioning.
Specifically, since the history $h$ is such that $\ssF_{i, t-1}$ did not occur, $P_{it}$ does not affect $\OPT$ in any way --- $i \in \ssI_t$ regardless of $P_{it}$. 
Similarly, $\Pr(P_{it} = 1 \;|\; \ssA_t = i, H_t = h, j \notin I_t \;\forall j > i) = p_i$.
Using this, we have that
\begin{align*}
\Pr(\ssA_t = i, A_t < i, H_t = h) 
&= (1-p_i)/p_i \Pr(\ssA_t = i, H_t = h, P_{it} = 1, j \notin I_t \;\forall j > i) \\
&= (1-p_i)/p_i \Pr(\ssA_t = i, H_t = h, A_t = i).
\end{align*}   

Then, we sum over all possible histories $h \in \cH_t$:
\begin{align*}
\Pr(\ssA_t = i, A_t < i)
&= \sum_{h \in \mathcal{H}_t} \Pr(\ssA_t = i, A_t < i, H_t = h) \\
&= \frac{1-p_i}{p_i} \sum_{h \in \mathcal{H}_t}\Pr(\ssA_t= A_t =i, H_t = h) \\
&\leq \frac{1-p_i}{p_i} \Pr(\ssA_t= A_t =i) \\
&\leq \frac{1-p}{p} \Pr(\ssA_t= A_t =i),
\end{align*}
where the first inequality is due to the fact that $\ssA_t= A_t =i$ can happen under a history $h \notin \mathcal{H}_t$, and the last inequality comes from the fact that $p_i \geq p$ for all $i$.

\subsection{Proof of \cref{thm:approx}}

Consider the first steps of \cref{theorem:weak} and \ref{theorem:main}, where we replace the $\GREEDY$ policy with $\ALG$. Specifically, we let $I_t$ be the resources that are available under $\ALG$ at time $t$, and $A_t$ to be the resource matched at time $t$ under $\ALG$.
Then, $O_{it}$ is the event that at time $t$, all resources $j \geq i$ are unavailable under $\ALG$.

We can then use the same decomposition of $\OPT$ as in \eqref{eq:decomp1}:
\begin{align} \label{eq:decomp2}
\OPT &= \bE\left[ \sumt \sum_{i \in [N]} r_i \bI( \ssA_t = i, \neg O_{it}) \right]
 +  \bE\left[ \sumt \sum_{i \in [N]} r_i \bI( \ssA_t = i, O_{it})\right]
\end{align}

Consider the first term in \eqref{eq:decomp2}.
The event $\{\ssA_t = i,\neg O_{it}\}$ implies that under $\ALG$, there exists at least one resource $j \geq i$ that is available, where $i$ is the resource matched by $\OPT$.
Therefore, by definition of $r^*(I_t)$, we have
\begin{align*}
\bE\left[ \sumt \sum_{i \in [N]} r_i \bI( \ssA_t = i, \neg O_{it}) \right] 
\leq 
\bE\left[ \sumt r^*(I_t) \right].
\end{align*}
By the approximation assumption on $\ALG$,
\begin{align*}
\bE\left[ \sumt r^*(I_t) \right]
\leq \frac{1}{\alpha}
\bE\left[ \sumt r_{A_t} \right] = \frac{1}{\alpha} \ALG.
\end{align*}
Therefore,
\begin{align} \label{eq:decomp3}
\OPT &\leq \frac{1}{\alpha} \ALG +  \bE\left[ \sumt \sum_{i \in [N]} r_i \bI( \ssA_t = i, O_{it})\right].
\end{align}  
The second term in \eqref{eq:decomp3} is defined as $\LOST$ in the proofs of \cref{theorem:weak} and \ref{theorem:main}.
Propositions~\ref{prop:bound_lost_weak} and \ref{prop:bound_lost} that upper bound $\LOST$ do not rely at all on the policy that is being compared to, and hence those results go through under $\ALG$.
Combining the Propositions with \eqref{eq:decomp3} yield the desired results of \cref{thm:approx}.

\section{Conclusion and Future Directions}\label{s.conclusion}

We study the performance of the $\GREEDY$ policy in the online bipartite matching problem with reusable resources. We show that $\GREEDY$ achieves a competitive ratio $\frac{1}{2-p}$ if every matched resource becomes available in the next period with at least probability $p$. Further, when the usage duration distributions are geometric with parameter $p$, we provide a stronger competitive ratio of $\frac{1+p}{2}$, which is proved to be tight. Moreover, we show that if an algorithm matches each request to a resource whose reward is $\alpha\in [0,1]$ of that in $\GREEDY$, then the competitive ratio of this algorithm under the two aforementioned assumptions are  $\frac{1}{1-p + 1/\alpha}$ and $\frac{1}{(1-p)/(1+p) + 1/\alpha }$ respectively. Our results indicate that the `reusability' of resources can significantly improve the performance of $\GREEDY$.

There are several questions for the future direction that this work opens up.
One direction is to understand whether the tight guarantee of $\frac{1+p}{2}$ can be shown for general usage distributions.
The current analysis that is used to show this tighter bound heavily relies on the geometric distribution, and hence new technical innovations may be needed to generalize the results.
Another direction is to investigate whether the performance bounds derived for $\GREEDY$ can be improved upon using other policies, using the same parameterization of the degree of reusability. There may be other natural policies that can take advantage of the reusability of resources to achieve better performance, and it would be interesting to explore these possibilities.

\bibliographystyle{informs2014} 
\bibliography{bibliography} 









\end{document}